\documentclass[notitlepage,aip,hidelinks,10pt,byrevtex,secnumarabic,twocolumn,nofootinbib,longbibliography,a4paper,showkeys,preprintnumbers]{revtex4-2}
\usepackage{graphicx}
\usepackage{amsmath}
\usepackage{amssymb}
\usepackage{float}
\usepackage{censor}
\usepackage{moreverb}
\usepackage{url}
\usepackage{tabularx,array}

\usepackage{xr-hyper}
\usepackage{xcolor}
\usepackage{colortbl}
\usepackage[draft,commandnameprefix=always]{changes}
\usepackage{tikz}
\usepackage{pgfplots}

\usetikzlibrary{shapes,arrows,arrows.meta,positioning,calc,external,patterns}
\usepgfplotslibrary{fillbetween}
\pgfplotsset{compat=1.11}

\usepackage[pdftex,
pdfauthor={Jose-Maria Martin-Olalla},
pdftitle={Thermal stability originates the vanishing of the specific heats at the absolute zero},
            pdfsubject={Foundations of Thermodynamics},
            pdfkeywords={second law of thermodynamics; third law of thermodynamics; specific heat; statistical mechanics; entropy; temperature; Carnot engine; foundations of thermodynamics; Carnot theorem; absolute zero},
            pdfproducer={Latex with hyperref, or other system},
            pdfcreator={pdflatex, or other tool}]{hyperref}

                        \newcommand\rorlink[2]{\href{https://ror.org/#2}{#1\,\includegraphics[scale=0.1]{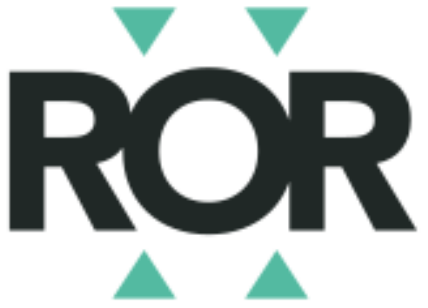}}
}
\usepackage{orcidlink}
\providecommand{\orcidlinki}[2]{\href{https://orcid.org/#2}{#1}\orcidlink{#2}}
            \usepackage{breakurl}
\usepackage{siunitx}
\usepackage{mathtools}
\usepackage[utf8]{inputenc}
\usepackage[english]{babel}
\usepackage{lineno}
\usepackage{array}

\newcommand{\zenodo}[1]{\doi{10.5281/zenodo.#1}}

\AtBeginDocument{\usepackage{booktabs}}

\graphicspath{{./figures/}}

\makeatletter\AtBeginDocument{\let\@elt\relax}\makeatother

\newcommand{\uaddress}{\rorlink{Universidad de Sevilla}{03yxnpp24}, Facultad de Física, Departamento de Física de la Materia Condensada,  ES41012 Sevilla, Spain}

\begin{document}
  \begin{widetext}
  This version is published in \texttt{arXiv}. It is a verbatim copy of \zenodo{17678316} and the version available at the Institutional Repository \burl{https://idus.us.es/handle/11441/179315}.

  Until the embargo period expires (3 December 2026) the Accepted Manuscript (\copyright 2025 IOP Publishing Ltd) is only available at Authors' home page \burl{https://personal.us.es/olalla}
  
  \textcolor{blue}{Una versión en lengua castellana está disponible en \zenodo{17678316}, en el repositorio institucional \burl{https://idus.us.es/handle/11441/179315}. Hasta que el periodo de embargo finalice (3 de diciembre del 2026) la traducción del Manuscrito Aceptado solo está disponible en la página personal \burl{https://personal.us.es/olalla}.}
    
  \end{widetext}

  \author{\orcidlinki{José-María Martín-Olalla}{0000-0002-3750-9113}}
\affiliation{\uaddress}
\email{olalla@us.es}

\title{Thermal stability originates the vanishing of the specific heats at the absolute zero}

\received[Posted ]{05 August 2025}
\revised{27 October 2025}
\accepted{21 November 2025}
\published{2 December 2025}

\preprint{\textcolor{blue}{This is Author's Original manuscript. Existe una versión en lengua castellana. The Version of Record}}
\preprint{\textcolor{blue}{is published at \emph{Physica Scripta} 2025 \textbf{100} 125206 \doi{10.1088/1402-4896/ae22a5} \copyright 2025 IOP Publishing Ltd. }}
\begin{abstract}
  The relationship between the vanishing of the specific heats as $T\to0^+$ and the thermal stability is discussed. The vanishing of the specific heats as fast as or faster than $T$ as $T\to0^+$ is the only consistent way by which states at the phase space boundary ($T=0$) can be thermally stable.

\end{abstract}


\maketitle

\section{Introduction}
\label{sec:introduction}

The general properties of matter in the neighborhood of $T=0$ are summarized by two independent observations: (i) the vanishing of the isothermal change of entropy (Nernst heat theorem) and (ii) the vanishing of the specific heats.\cite{Nernst1918,Nernst1924} The two observations traditionally form the third law of thermodynamics, as they could not be deduced from the first and second laws of thermodynamics.\cite{pippard-1957,hatsopoulos-1965,kestin-1968ii,callen-85} The relationship between these two observations and the unattainability of the absolute zero has sparked discussion in the past.\cite{Yan1988,Landsberg1989,Oppenheim1989,Yan2001,Martin-Olalla2003b,Su2022,Martin-Olalla2024f,Martin-Olalla2024gi}

Recently, it was assumed that, within the framework of the second law, $T=0$ must be determined by a Carnot thermometer. Therefore, Einstein's objection that a Carnot engine at $T=0$ cannot operate ``in practice''\cite{Solvay-1913,Kox2006} was rebutted, and the Nernst theorem probed.\cite{Martin-Olalla2025c} This leaves the vanishing of the specific heats as the only general property of matter in the neighborhood of $T=0$ that cannot be predicted by the first and second law.

Traditionally, the vanishing of the specific heats as $T\to0^+$ has been associated with the definiteness of Clausius' entropy at $T=0$.\cite{Planck-1911} Alternatively, the vanishing of the specific heats is strong evidence of the validity of Boltzmann's entropy, since the number of accessible microstates $\Omega$ at $T=0$ cannot vanish in a real system and, therefore, $S\sim\log\Omega$ must remain finite. Finally, classical models such as the ideal gas or the crystalline solid fail to predict the vanishing of the specific heats.\cite{huang-87} Only when quantum mechanics is incorporated do these models satisfy the requirement.\cite{Einstein1907} Therefore, the origin of this requirement is tipically linked to quantum physics.\cite{Wheeler1991,Masanes2017}

The goal of this manuscript is to present a thermodynamic argument that explains why real systems exhibit vanishing specific heats as $T\to0^+$ at least as faster as $T$. The argument relies solely on the thermal stability of the states in the phase space boundary, which is the condition $T(S,X)=0$,\cite{Landsberg1956} where $S$ is the entropy and $X$ is a suitable mechanical parameter such as volume or magnetization. If successful, every aspect related to the third law of thermodynamics would then have been derived from the framework of the first and second law of thermodynamics.

\section{Methods}
\label{sec:framework-methods}

Thermodynamic equilibrium is a spatially homogeneous state. Its stability means that every spatial inhomogeneity arising from an equilibrium state will initiate a process that mitigates the inhomogeneity, restoring the original state.

With equilibrium states determined by minimizing of the energy $U$ (or maximizing the entropy) for a given entropy (or energy), stability conditions are associated with:
\begin{equation}
  \label{eq:4}
  U(S_e+\delta S,X_e+\delta X)+U(S_e-\delta S,X_e-\delta X)>2U(S_e,X_e),
\end{equation}
where $(S_e,X_e)$ is the state under study and $(\delta S,\delta X)$ drives the inhomogeneity. Condition~(\ref{eq:4}) means $U(S,X)$ must be a convex function of $(S,X)$. Physically, this implies that $U(S,X)$ increases with the inhomogeneity and, therefore, it can be followed by a regular process that, by decreasing $U(S,X)$, restores back $U(S_e,X_e)$. 

Sufficient conditions for~(\ref{eq:4}) are:\cite{callen-85}
\begin{equation}
  \label{eq:3}
U_{ss}>0; U_{xx}>0; \det\mathcal{H}_u>0, 
\end{equation}
where $U_{ii}$ stands for $(\partial^2U/\partial i^2)$ and $\det\mathcal{H}_u$ is the determinant of the Hessian matrix. These conditions are obtained by expanding $U(S,X)$ and analysing the leading term, which is quadratic in $(\delta S,\delta X)$.

The above analysis becomes problematic at the phase space boundary ($T(S,X)=0$). After the minimum entropy $S_0$ is reached at the boundary ---the Nernst theorem is assumed to be valid\cite{Martin-Olalla2025c}---, an inhomogeneity such as $S_0-\delta S$ seems not possible. However, the principle of continuity suggests this analysis remains valid in the neighborhood of $T\to0^+$.

This study focuses solely in the thermal stability, associated with entropy inhomogeneities that develops at constant mechanical parameter $X$ ($\delta X=0$) or its conjugate $Y$ ($\delta Y=0$). They can be expressed from~(\ref{eq:3}) as:
\begin{subequations}
\label{eq:1g}  
\begin{align}
  \label{eq:1}
  U_{ss}=\left(\frac{\partial^2U}{\partial S^2}\right)_x&=\left(\frac{\partial T}{\partial S}\right)_x=\frac{T}{C_x}>0,\\
  \label{eq:2}
  H_{ss}=\left(\frac{\partial^2H}{\partial S^2}\right)_y&=\left(\frac{\partial T}{\partial S}\right)_y=\frac{T}{C_y}>0.
\end{align}
\end{subequations}
where $C_i$ is the heat capacity at $i$ constant and $H(S,Y)$ is the enthalpy. It must be noted that $\det\mathcal{H}_u=U_{xx}\times H_{ss}$. Condition~(\ref{eq:2}) directly conveys the information associated with the requirement $\det\mathcal{H}_u>0$. As an example, ordinary first-order phase transitions in hydrostatic systems ---a paradigm of stability loss--- occur when $H_{ss}$ vanishes due to $C_y\to\infty$.

\section{Results}
\label{sec:results}

Conditions~(\ref{eq:1g}) can be analyzed in the limit of $T\to0^+$ for a leading analytical term $C_i\sim T^b$, where $b\geq0$ is a constant.\footnote{This choice excludes specific heats that vanish non-analytically  as $T\to0^+$, such as $C_i\sim -1/\log(T)$.\cite{Mattis2003,Liu2025}} The following results arise. First, for $b\geq1$, the heat capacities vanish at a rate at least as fast as $T$. In this case, the equilibrium states at the phase space boundary ($T=0$) are thermally stable, as conditions~(\ref{eq:1g}) upheld as $T\to0^+$. The entropy at $T=0$ would remain finite.

Second, for $b\in(0,1)$, the heat capacities vanish at a slower rate than $T$. Consequently, $U_{ss}\to0^+$ and $H_{ss}\to0^+$ as $T\to0^+$. This indicates that the equilibrium states at the phase space boundary ($T=0$) would be thermally unstable even though $S(T,X)$ would still remain finite at $T=0$.

Third, for $b=0$ the heat capacities do not vanish and neither $S(T,X)$ is finite at $T=0$ nor thermal stability is sustained at the boundary of the phase space.

Experimental observations in real systems and useful models, such as the quantum ideal gas or the Debye model, are consistent with $b\geq1$ in the phase space boundary ($T=0$). This is noted when discussing the third law.\cite{callen-85,huang-87,ashcroft-1976}

\begin{figure*}[t]

  \includegraphics[width=0.8\textwidth]{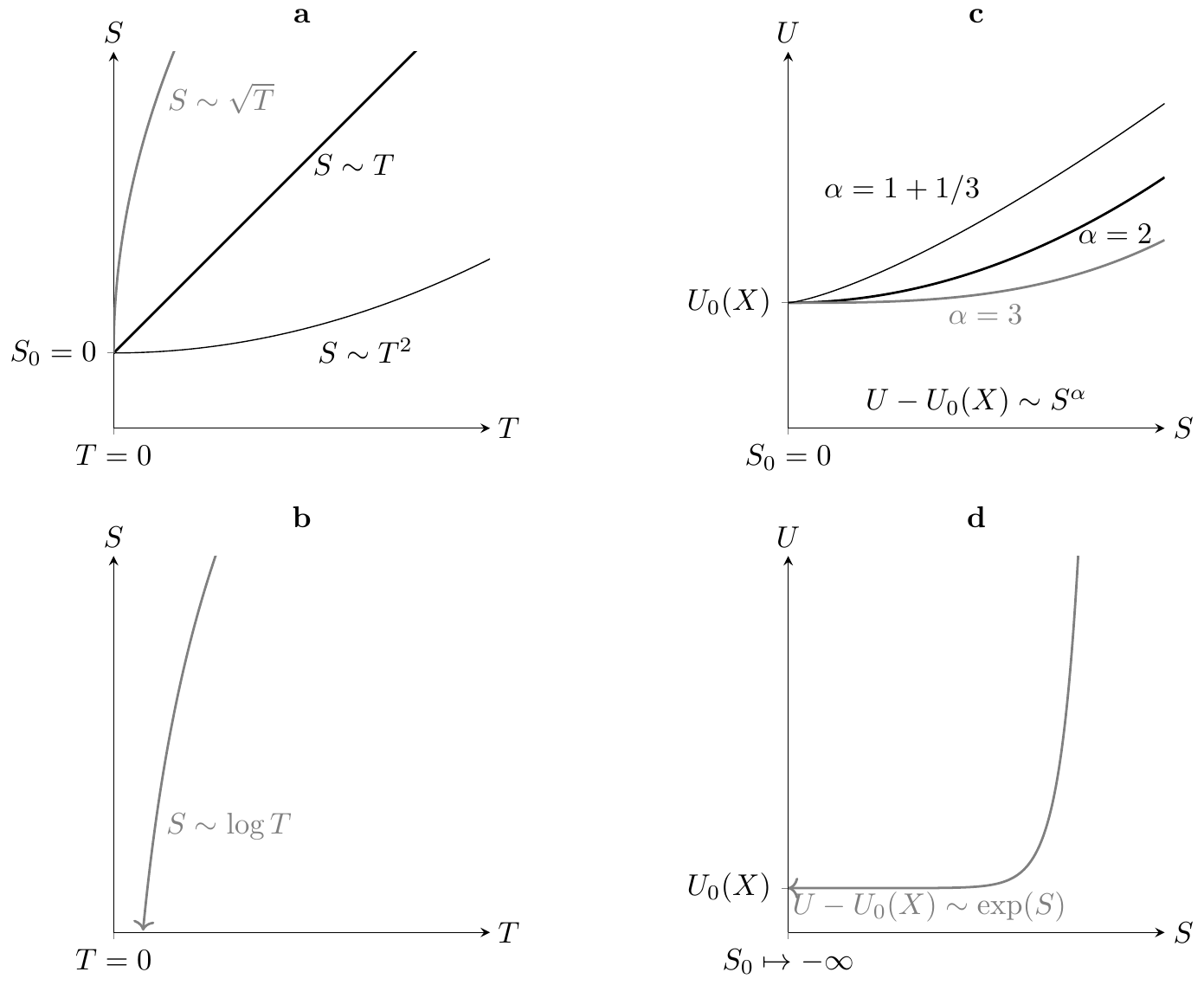}

  \caption{Panel \textbf{a}, the $S-T$ chart for $C_x\sim T^b$ for $b=1/2$ (gray), $b=1$ (dark, thick) and $b=2$ (dark, thin). All three cases yield a finite $S_0$ but $b=1/2$ is unstable due to the diverging slope at $T=0$. Panel \textbf{b}, as panel \textbf{a} but for the case $b=0$ which translates in a $\log T$ dependence. This is unstable and unbounded. Panel \textbf{c},  the $U-S$ chart for the cases displayed in panel \textbf{a}. Stability requires a stiff relationship between $U$ and $S$, with a parabola (dark thick line) as the critical condition. The case $b=1/2$ (gray line) yields a very shallow flat arrival at $U_0(X)$ and the equilibrium state is unstable. Panel \textbf{d}, the same as panel \textbf{c} but for the case $b=0$. Here $U(S,X)$ is too flat to sustain stability, and the domain of $S$ is unbounded.}
  \label{fig:uno}
\end{figure*}

The above analysis can be visualized in the $S-T$ plane with the help of iso-$X$ lines. Conditions~(\ref{eq:1g}) indicate that for a system to maintain thermal stability at the phase space boundary, its iso-$X$ lines and its iso-$Y$ lines must have a finite, positive slope $(\partial S/\partial T)_i$. This ensures a finite, well-defined entropy at the intercept with $T=0$. In figure~\ref{fig:uno}a the thick, dark line shows $b=1$ and the thin, dark line $b=2$. The case $b<1$ is noted by $b=1/2$ or $S\sim\sqrt{T}$ (gray line). Now the slope becomes indefinitely large as $T\to0^+$, and thermal stability is not achieved at the phase space boundary $T=0$. Finally, if the heat capacity is a non-zero constant (case $b=0$), then $S\sim\log T$ (see figure~\ref{fig:uno}b). In this scenario neither $S_0$ is bounded nor stability is guaranteed at the phase space boundary.

In the $U-S$ plane (or in the $H-S$ plane), $C_i\sim T^b$ translates into  a functional form $U(S,X)- U_0(X) \sim S^\alpha$, with $\alpha=1+1/b$ and $b>0$. This relationship shows that thermal stability requires $\alpha\leq2$. Figure~\ref{fig:uno}c illustrates this scenario where the thick, dark represents the critical condition for stability ($\alpha=2$, corresponding to $b=1$). Conversely, an exponent $\alpha>2$ (meaning $b<1$) would result an unstable phase space boundary, as shown by the gray line in figure~\ref{fig:uno}c.

It is important to note that $U(S,X)$ is always flat at the phase space boundary because $(\partial U/\partial S)_x=0$ sets the phase space boundary. Also $U(S,X)$ is convex as required by~(\ref{eq:4}). The sufficient conditions for stability~(\ref{eq:3}) are achieved only if the curvature of $U(S,X)$ is at least as stiff as a parabola ($\alpha=2$). If the curvature is too shallow, as when $\alpha>2$, then the system cannot sustain stability at the phase space boundary. This behaviour is exacerbated as $\alpha\to\infty$ ($b\to0$), which corresponds to $U(S,X)-U_0(X)\sim\exp(S)$, see figure~\ref{fig:uno}d.

\section{Discussion}
\label{sec:discussion}

The vanishing of the specific heats as $T\to0^+$ plays a singular role among the empirical laws of thermodynamics. A negation of the first law or a negation of the second law makes a global impact since a perpetuum mobile of the first kind, or a perpetuum mobile of the second kind would be possible. A negation of the Nernst heat statement also makes a global impact since $T=0$ would be accessible, thus enlarging the domain of accessed temperatures. In contrast, the impact of a would-be real system whose specific heat does not vanish as $T\to0^+$ is limited to this specific system. The same can be said for a system whose entropy is not finite at $T=0$.

In addition, while the vanishing of specific heats as $T\to0^+$ is a valid general observation, it fails to exclude scenarios that are never observed in nature and are characterized solely by their lack of thermal stability. This includes analytical cases such as $C_i\sim T^b$, with $b\in(0,1)$ (as an example $C_i\sim\sqrt{T}$), and non-analytical cases such as $C_i\sim -1/\log(T)$.\cite{Mattis2003} The sufficient conditions for stability, derived from the second law of thermodynamics, impose a stricter requirement on the system's behavior that is noteworthy to highlight: heat capacities must vanish at a rate at least as fast as $T$.

\section{Conclusion}
\label{sec:conclusion}

The experimental evidence regarding the behavior of specific heats as $T\to0^+$ is most accurately and comprehensively codified by the statement that the sufficient conditions for thermal stability are sustained as $T\to0^+$.

This direct link between stability and the observed behavior of specific heats resolves a conceptual mismatch in traditional discussions. Consequently, the vanishing of specific heats at a rate at least as fast as $T$ should be viewed not as a new, independent law, but as a confirmatory result of the existing framework of thermodynamics. The evidence confirms that real systems maintain thermal stability down to absolute zero, a principle already inherent in the first and second laws.

%

\section*{Timeline}
\label{sec:timeline}

This work was conceptualized around June 30th while preparing a nice talk/conversation with \href{https://www.youtube.com/@EnriqueFBorja}{\texttt{Cuentos Cuánticos} (Quantum Stories)} hosted by Enrique Fernández Borja on the occasion of Ref.~[\onlinecite{Martin-Olalla2025c}]. The talk ---in Spanish language and two-hour long--- is available on \href{https://youtu.be/piz9v9Rrk98?si=-N7DkpHbpnhLbi84\&t=4807}{https://youtu.be/piz9v9Rrk98?si=-N7DkpHbpnhLbi84\&t=4807}. The link points to the discussion on the stability. See also the subsequent comments for refinements.

The manuscript was started, drafted and posted on August 05 2025, on the feast of Our Lady of the Snows. On August 07 2025, a new version was posted, which was updated on August 09 2025, and on August 15 2025.

\end{document}